\documentstyle[12pt]{article}

\begin{document}

\title{LHC  $Z^{'}$ discovery potential  for models with 
continuously distributed mass} 

\author{N.V.Krasnikov  
  \\
INR RAN, Moscow 117312}

\maketitle
\begin{abstract}
We study   the Large Hadron Collider (LHC) discovery potential 
for $Z^{`}$ models with continuously distributed mass for 
$\sqrt{s} = 7, ~10$ and $14~TeV$ centre-of-mass energies.
One of  possible LHC signatures for such models is the existence of   
broad  resonance structure in Drell-Yan reaction 
$pp \rightarrow Z^{'} + ... \rightarrow l^{+}l^{-} + ...$. 
\end{abstract}

\newpage

The search for new vector $Z^`$ bosons \cite{1} is one of the most 
important goals  of  the Large Hadron Collider (LHC). 
The LHC $Z^`$ boson discovery potential at CMS and ATLAS detectors has been 
studied for narrow $Z^`$ resonances in refs.\cite{2,3}. 
The $Z^`$ models with broad resonance structure naturally 
arise in the context of 
$Z^{`}$ models with continuously distributed masses \cite{4,5}.
Note that recent notion of  unparticle   
introduced by Georgi \cite{6} (unparticle phenomenology 
is discussed in  refs.\cite{7}) 
can be interpreted 
as a 
particular case of a field  with continuously 
distributed mass \cite{4,5,8,9,10,11}.  Also scenario with large 
invisible 
decay width can be realized if  $Z^`$ boson couples to 
neutral leptons and   decays mainly into heavy neutral leptons. 

In this paper  we present the  LHC discovery potential 
for the $Z^{`}$ models with continuously distributed mass for 
$\sqrt{s} = 7, ~10$ and $14~TeV$ centre-of-mass energies.
\footnote{In recent papers \cite{12} the LHC $Z^{`}$ boson discovery 
potential was studied 
for    energies $\sqrt{s} = 7, ~10$ and $14~TeV$.}
 
Consider the Stueckelberg Lagrangian \cite{13} \begin{equation} L_0 = 
\sum_{k=1}^{N}[-\frac{1}{4}F^{\mu\nu,k}F_{\mu\nu,k} 
+\frac{m^2_k}{2}(A_{\mu,k} - \partial_{\mu}\phi_k)^2] \,, \end{equation} 
where $F_{\mu\nu,k} = \partial_{\mu}A_{\nu,k} - \partial_{\nu}A_{\mu,k}$. 
The Lagrangian (1) is invariant under gauge transformations 
\begin{equation} 
A_{\mu,k} \rightarrow A_{\mu,k} +\partial_{\mu}\alpha_k 
\,, \end{equation} \begin{equation} \phi_k \rightarrow \phi_k + \alpha_k \, 
\end{equation} 
and it describes N free massive vector fields with 
masses $m_k$. For the field $B_{\mu} = \sum_{k=1}^N c_kA_{\mu,k}$ the 
propagator in transverse gauge is 
\begin{equation} 
D_{\mu\nu}(p) = 
(g_{\mu\nu} - \frac{p_{\mu}p_{\nu}}{p^2}) (\sum^N_{k=1} 
\frac{|c|^2_k}{p^2-m^2_k}) \,. 
\end{equation} 
In the limit $N \rightarrow 
\infty $ 
\begin{equation} 
D_{\mu\nu}(p)  \rightarrow (g_{\mu\nu} - 
\frac{p_{\mu}p_{\nu}}{p^2}) D_{int}(p^2)\,, 
\end{equation} 
where 
\begin{equation} 
D_{int}(p^2)  = \int _0^{\infty}\frac{\rho(t)}{p^2 - t 
+i\epsilon}dt 
\end{equation} 
and $\rho(t) = lim_{N\rightarrow \infty 
}|c^2_k|\delta(t -m^2_k)  \geq 0$.

The fields with continuously distributed masses arise naturally 
in d-dimensional field theories \cite{4}. 
Consider the 5 -dimensional extension of the Stueckelberg 
Lagrangian (1), namely:
\begin{equation}
L_5 = -\frac{1}{4}F^{\mu\nu}(x, x_4)
F_{\mu\nu}(x ,x_4) +\frac{1}{2}((A_{\mu}(x ,x_4) - 
\partial_{\mu}\Phi(x ,x_4))f(-\partial_4^2)(A^{\mu}(x ,x_4) - 
\partial^{\mu}\Phi(x ,x_4)) \,,
\end{equation} 
where $F_{\mu\nu}(x ,x_4) = \partial_{\mu}A_{\nu}(x ,x_4)
- \partial_{\nu}A_{\mu}(x ,x_4)$, $\partial_4 = 
\frac{\partial}{\partial x_4}$, $x = (x^0,x^1,x^2,x^3)$; $\mu, \nu = 0,1,2,3$. 
The Lagrangian (7) is invariant under gauge transformations
\begin{equation}
A_{\mu}(x ,x_4) \rightarrow A_{\mu}(x ,x_4) +
\partial_{\mu}\alpha(x ,x_4)\,,
\end{equation}
\begin{equation}
\Phi(x ,x_4) \rightarrow \Phi(x ,x_4) + \alpha(x ,x_4)\,.
\end{equation}
Note that the Lagrangian (7) for arbitrary 
function $f(-\partial^2_4)$    is invariant only under 
four-dimensional Poincare group.
For the Lagrangian (7) the propagator for the vector field 
$A_{\mu}(x ,x_4)$ in transverse gauge is 
\begin{equation}
D_{\mu\nu}(p ,p_4) = (g_{\mu\nu} - 
\frac{p_{\mu}p_{\nu}}{p^2})  \frac{1}{p^2 - 
f(p^2_4)   +i\epsilon}\,,
\end{equation}
where $p^2 \equiv p^{\mu}p_{\mu}$.
The propagator for the four-dimensional vector field 
$A_{\mu}(x, x_4 =0)$ in transverse gauge has the form
\begin{equation}
D_{\mu\nu}(p) = (g_{\mu\nu} - 
\frac{p_{\mu}p_{\nu}}{p^2})
\frac{1}{2\pi}\int_{-\infty}^{+\infty}dp_4
\frac{1}{p^2 - 
f(p^2_4)+i\epsilon} = 
\int_{0}^{\infty}dt\frac{\rho(t)}{p^{\mu}p_{\mu} -t +i\epsilon}\,,
\end{equation}
where
\begin{equation}
\rho(t) = \frac{1}{\pi} \frac{d\sqrt{f^{(-1)}(t)}}{dt}\,,
\end{equation}
where $f(f^{-1}(t)) = t $.

Consider the interaction of the 
five-dimensional field $A_{\mu}(x ,x_4)$ with 
the four-dimensional fermion field $\psi(x)$ in the form
\begin{equation}
L_{int} = g\bar{\psi}(x)\gamma^{\mu}\psi(x)
A_{\mu}(x ,x_4 = 0) \,.
\end{equation}
In our model vector field $A_{\mu}(x ,x_4)$ lives 
in five-dimensional space-time whereas other matter fields 
(quarks, leptons ..) live in standard five-dimensional field. 
\footnote{The model (13) is similar to ADD model \cite{14} in 
which gravity lives in $(d>4$) space-time whereas other 
matter fields live in four-dimensional space-time.}  
The Feynman rules for the model (13) coincide with standard 
Feynman rules for QED except the replacement of photon propagator 
\begin{equation}
\frac{1}{p^2 +i\epsilon} \rightarrow \int^{+\infty}_{0}
\frac{\rho(t)dt}{p^2  - t +  i\epsilon} \,,
\end{equation}
Note that considered model as the uncompactified ADD model is 
unitary only in 5-dimensional space-time since vector fields 
$A_{\mu}(x, x_4)$ propagates in 5-dimensional space-time.

As an example we consider the  spectral density $\rho(t)$ 
of the Breit-Wigner type.
The following approximate relation takes place:
\begin{equation}
\frac{1}{p^2 -m^2 +im\Gamma} \approx
\int_{0}^{\infty} \frac{\rho(t)dt}{p^2-t+i\epsilon}\,,
\end{equation}
where 
\begin{equation}
\rho(t) = \frac{1}{\pi}\frac{\Gamma m}{(t-m^2)^2 + \Gamma^2m^2}\,.
\end{equation}
 As it follows from the equalities (15,16) we can consider 
vector particle with 
continuously distributed mass as a particle with some 
internal decay width $\Gamma$ into additional dimension.
\footnote{Note that in RS 2 model \cite{15} 
massive particles initially located on our brane may leave the 
brane and disappear into extra dimensions  \cite {16}.}
 
In most standard $Z^`$ models 
\cite{1}
the total decay width $\Gamma_{tot}$ is rather narrow. 
Typically $\Gamma_{tot} \leq  O(10^{-2})\cdot m_{Z^{`}}$.
It should be noted that the  interaction of $Z^{`}$  boson with additional 
neutral fermions 
increases invisible  $Z^`$ decay width and can lead to 
large  total decay width of the $Z^{`}$ boson.
One can introduce the interaction of the field $B_{\mu}$ with  
quark and lepton fields in a standard way as
 \begin{equation}
L_{int} = J_{\mu}  B^{\mu} \,,
\end{equation}
where 
\begin{equation}
J_{\mu} = \sum_k [g_{kL}\bar{q}_L\gamma_{\mu}q_L + 
g_{kR} \bar{q}_R\gamma_{\mu}q_R +
g_{lL}\bar{l}_L\gamma_{\mu}l_L +
g_{lR}\bar{l}_R\gamma_{\mu}l_{R}]\,.
\end{equation}

The Feynman rules for this model coincide with Feynman rules for 
standard $Z^`$ model  except  the replacement of the standard vector 
$Z^`$ boson propagator

\begin{equation}
D^{tr}_ {\mu\nu}(p) = 
(g_{\mu\nu} - \frac{p_{\mu}p_{\nu}}{p^2})\frac{1}{p^2 - m^2_{Z^`}} \rightarrow 
(g_{\mu\nu} - \frac{p_{\mu}p_{\nu}}{p^2})D_{int}(p^2) \,,
\end{equation}
\begin{equation}
D_{int}(p^2)  =   \int_{0}^{\infty} \frac{\rho(t)dt}{p^2-t+i\epsilon}\,,
\end{equation}

This generalization 
preserves the renormalizability for finite  
$ \int _0^{\infty}\rho(t)dt$ because the ultraviolet asymptotic of 
$ D_{int}(p^2)$ coincides with  free $Z^{`}$ boson
 propagator asymptotic $\frac{1}{p^2}$
\footnote{Note that for $\rho(t) \sim t^{\delta -1}$ we 
reproduce the case of vector 
unparticle with propagator $\sim\frac{1}{(p^2)^{1-\delta}}$.}.

In this paper we consider three $Z'$ models. In model A the 
$Z^`$ boson  interaction with quarks and leptons is the same as 
for standard $Z$ boson. In model B the interaction of $Z^`$ boson 
with quark and lepton fields is determined by the current
\begin{equation}
J_{\mu} = \frac{g}{2}  [\sum_{q,l}(\bar{q}\gamma_{\mu}(1 + \gamma_5)q + 
\bar{l}\gamma_{\mu}(1 + \gamma_5)l ]\,,
\end{equation}
where $g  = 0.65  $ is the Standard Model $SU(2)_L$ coupling constant.
\footnote{Model B has $\gamma_5$-anomaly. By the introduction of 
additional neutral fermions it is possible to get rid of 
$\gamma_5$-anomalies. At any rate we consider model B as 
a toy model.}
In model C the $Z^`$ boson interacts with (B-L) current 
\begin{equation}
L_{mu} = g_{B-L}\sum_{q,l}[ \frac{1}{3} \bar{q}\gamma_{\mu}q - 
\bar{l}\gamma_{\mu}l]
\end{equation}
with coupling constant $g_{B-L} = 0.9$. 

For standard $Z^`$ boson the matrix element of 
the Drell-Yan reaction at quark-parton level  $q\bar{q} 
\rightarrow Z^`  \rightarrow l^+l^-$ is proportional to

\begin{equation}
M(q \bar{q} \rightarrow Z^` \rightarrow l^+l^- ) \sim 
\frac{1}{s -M^2_{Z^`} +i\Gamma_{tot}M_{Z^`}}\,,
\end{equation} 
where $\Gamma_{tot}$ is 
total decay width of the $Z^`$ boson. For our model with free vector 
propagator (7,8) the matrix element 
is proportional to 
\begin{equation}
M(q \bar{q} \rightarrow Z^` \rightarrow l^+l^-) \sim
\frac{1}{s - 
M^2_{Z^`} + i\Gamma_{tot}M_{Z^{`}} + i\Gamma M_{Z^`}}\,. 
\end{equation} 
The difference between our model and standard case consists in the 
replacement 
$\Gamma_{tot} \rightarrow \Gamma_{tot} + \Gamma $. 
It means that our model is equivalent to standard $Z^`$ model 
with additional invisible decay width $\Gamma$. This fact allows 
to use standard simulation code PYTHIA \cite{17}. 
To make the invisible
$Z^{`}$ resonance decay width $\Gamma$ at the PYTHIA level 
we modified the $Z^`$ boson 
interaction with the 
first flavor of neutrino, namely we made the replacement
 
$$Z_{\mu}^{`}g_{\nu} \bar{\nu}(\gamma_{\mu} +
\gamma_{\mu}\gamma_5)\nu   \rightarrow
Z_{\mu}^{`}g_{\nu} \bar{\nu}(\gamma_{\mu}(1 + u) 
+\gamma_{\mu}\gamma_5 (1 -u)) \nu, $$
where $u$ is some parameter. The parameter  $u =0$ corresponds to the case 
of standard interaction 
and the invisible decay width into right-handed neutrino 
is proportional to $|u|^2$ and for large
$u$ invisible decay width  dominates. In other words, 
we introduced additional interaction of 
the $Z^`$ boson with right-handed neutrino that allows to make 
the $Z^`$ boson rather wide. 
\footnote{For large parameter $u$ the coupling constant $g_{\nu_R}u$  of 
the interaction $Z^`$ boson with right-handed neutrino is large that 
 leads to the existence  of Landau pole singularity 
for the effective charge at  TeV  energies 
and hence to the breakdown of unitarity at least within 
perturbation theory. However in the 
$Z^`$ models with 
continuously distributed mass we don't have these problems. We introduce 
the artificial interaction with right-handed neutrino in order 
to obtain the  large $Z^`$ 
invisible decay width at the PYTHIA level.}   
The   values of total decay width 
$\Gamma_{tot}$ for some parameters $u$ for 
models A, B and C are given in Table 1.

The reaction we are interested in is di-lepton production
$$pp \rightarrow \gamma^* , Z^*, Z^{`*} + ... 
\rightarrow l^+l^-  +...$$
  Here $ l =e,\mu$. The main 
background is the  Drell-Yan production
$PP \rightarrow \gamma^*, Z^*   + ... 
\rightarrow l^+l^- + ...$. Other backgrounds like $WW, ZZ, WZ$ or 
 $t\bar{t}$ are small. For leptons standard acceptance cuts are the following:
\begin{equation}
p^l_T  > 10~GeV \,,
\end{equation}
\begin{equation}
|\eta^l| < 2.4 \,.
\end{equation}
For  di-lepton invariant mass $M_{ll}$ we apply cut
\begin{equation}
M_{ll} > M_{Z^`}\,.
\end{equation}
This cut differs from standard cut
\begin{equation}
|M_{ll} - M_{Z`}| < \frac{\Gamma_{Z^`}}{2}
\end{equation}
often used for the estimation of the LHC $Z^`$ boson discovery potential 
for standard case of narrow $Z^`$ boson. The reason is that the Drell-Yan 
cross section is sharply decreasing function on 
$M_{ll}$ and for wide $Z^`$ boson the 
Drell-Yan background dominates in
mass region
  $M_{Z^`} - \frac{\Gamma_{Z^`}}{2} < M_{ll} < M_{Z^`}$.   
In our calculations we use PYTHIA 6.3 code \cite{17}   with STEQ6L 
\cite{18} parton distribution functions evaluated at the scale 
$Q^2 = M^2_{Z^`}$. We did not take into account 
K-factors. An account of K-factor which is similar for both background 
and signal leads to the increase of significance. Also we don't take 
into account detector effects like effectiveness of electron or 
muon registration but at the CMS and ATLAS detectors the effectiveness of 
lepton 
registration is typically greater  than 80 percent \cite{2,3} that can 
decrease 
significance by factor $\geq 0.8$. An account of nonzero K-factor 
partly compensates this decrease. For the LHC total energy
$\sqrt{s} = 14~TeV$ and for $Z^`$ boson masses $1~TeV \leq M_{Z^{`}} 
\leq 4~TeV$ K-factor for both signal and 
background is approximately 1.3 - 1.4 \cite{2}. An account of 
K-factor leads to the increase of the significance by factor $\sqrt{K} 
\approx 1.1 - 1.2$.     
For the estimation of the significance   
we use the method \cite{19} based on frequentist approach 
which for the case of large statistics ($N_S \gg 1, N_B \gg 1$) 
leads to the approximate formula 
$S = 2(\sqrt{N_S +N_B} - \sqrt{N_B})$ for significance.

The first two years LHC will work with center-of-mass energy 
$\sqrt{s} = 7~TeV$ and with total integrated luminosity 
$L_{tot} = O(1)~fb^{-1}$. 
Finally  the energy will be increased up to $\sqrt{s} = 14~TeV$. 
In our estimates we use total integrated luminosity $L_{tot} = 1~fb^{-1}$ for 
energies $\sqrt{s} = 7~TeV, ~10~TeV$ and $L_{tot} = 100~fb^{-1}$
for the  LHC energy $\sqrt{s} = 14~TeV$. For models A, B and C we 
studied the  $Z^`$ bosons with total decay widths  shown in Table 1. 
Our results are summarized in Tables 2-10.

Table 1. The ratio $\frac{\Gamma_{Z^`,tot}}{M_{Z^`}}$ for 
models A,B,C and parameters $u =0,3,5,10,15,20$. 
\begin{center}
\begin{tabular}{|l| |l| |l| |l|}
\hline
 u-parameter  &$\frac{\Gamma_{Z^`,tot}}{M_{Z~}}$(mod.A)   
& $\frac{\Gamma_{Z^`,tot}}{M_{Z~}}$(mod.B) & $\frac{\Gamma_{Z^`,tot}}{M_{Z~}}$(mod.C) \\
\hline
 $u =0$  & 0.031  &  0.046    & $6.2\cdot10^{-3}$ \\
\hline
$u =3$   &  0.048  & 0.062  & 0.028  \\
\hline
$u =5$  & 0.078  &  0.092    & 0.061 \\
\hline
$u = 10$   & 0.22  & 0.23  & 0.21  \\
\hline 
$u  = 15$    & 0.45 &  0.47   &  0.45 \\
\hline
$u = 20$    & 0.78  &  0.79  &  0.79 \\

\hline
\end{tabular}
\end{center} 

Table  2. The LHC cross sections for signal plus background in fb and significances (in brackets)  
for total energy $\sqrt{s} = 7~TeV$ and $M_{Z^`} = 1~TeV$. The background cross section is $\sigma_B = 1.9~fb$ 
and total integrated luminosity is $L_{tot} = 1~fb^{-1}$. 
\begin{center}
\begin{tabular}{|l| |l| |l| |l|}
\hline
 u-parameter  & model A   & model B & model C \\
\hline
 $u =0$  & 94 (17)  &  166 (23)    &  24 (7.3) \\
\hline
$u =3$   &  58 (13)  & 114 (18)  & 5.8 (2.7)  \\
\hline
$u =5$  & 36 (6.2)  &  72 (14)    & 3.2 (1.7) \\
\hline
$u = 10$   & 12 (4.7)  & 20 (6.5)  & non detectable  \\
\hline 
$u  = 15$    & 5.6 (2.7) &  7.2 (3.2)   &  non detectable \\
\hline
$u = 20$    & 3.6 (1.8)  &  5.2 (2.5)  & non detectable  \\

\hline
\end{tabular}
\end{center} 

Table  3. The LHC cross sections for signal plus background in fb and significances (in brackets)  
for total energy $\sqrt{s} = 7~TeV$ and $M_{Z^`} = 1.2~TeV$. The background cross section is $\sigma_B = 0.68~fb$ 
and total integrated luminosity is $L_{tot} = 1~fb^{-1}$. 
\begin{center}
\begin{tabular}{|l| |l| |l| |l|}
\hline
 u-parameter  & model A   & model B & model C \\
\hline
 $u =0$  & 47 (12)  &  83 (16)    &  12 (5.2) \\
\hline
$u =3$   &  29 (9.0)  & 57 (13)  & 2.9 (1.9)  \\
\hline
$u =5$  & 18 (6.2)  &  36 (10)    & 1.6 (1.2) \\
\hline
$u = 10$   & 6 (3.3)  & 10 (4.6)  & non detectable  \\
\hline 
$u  = 15$    & 2.8 (1.9) &  3.6 (2.3)   &  non detectable \\
\hline
$u = 20$    & 1.8 (1.3)  &  2.6 (1.8)  & non detectable  \\

\hline
\end{tabular}
\end{center} 

Table  4. The LHC cross sections for signal plus background in fb and significances (in brackets)  
for total energy $\sqrt{s} = 10~TeV$ and $M_{Z^`} = 1.2~TeV$. The 
background cross section is $\sigma_B = 2.2~fb$ 
and total integrated luminosity is $L_{tot} = 1~fb^{-1}$. 
\begin{center}
\begin{tabular}{|l| |l| |l| |l|}
\hline
 u-parameter  & model A   & model B & model C \\
\hline
 $u =0$  & 102 (17)  &  176 (24)    &  26 (7.3) \\
\hline
$u =3$   &  66 (13)  & 124 (19)  & 6.4 (2.8)  \\
\hline
$u =5$  & 40 (10)  &  81 (15)    & 3.6 (1.8) \\
\hline
$u = 10$   & 13.6 (4.9)  & 24 (7.1)  & non detectable  \\
\hline 
$u  = 15$    & 6.6 (3.0) &  8.4 (3.5)   &  non detectable \\
\hline
$u = 20$    & 4 (2.0)  &  4.1 (2.0)  & non detectable  \\

\hline
\end{tabular}
\end{center}

Table  5. The LHC cross sections for signal plus background in fb and significances (in brackets)  
for total energy $\sqrt{s} = 10~TeV$ and $M_{Z^`} = 1.5~TeV$. The background cross section is $\sigma_B = 0.66~fb$ 
and total integrated luminosity is $L_{tot} = 1~fb^{-1}$. 
\begin{center}
\begin{tabular}{|l| |l| |l| |l|}
\hline
 u-parameter  & model A   & model B & model C \\
\hline
 $u =0$  & 34 (10)  &  58 (14)    &  8.8 (4.5) \\
\hline
$u =3$   &  22 (7.8)  & 40 (11)  & 2.1 (1.6)  \\
\hline
$u =5$  & 13 (5.6)  &  26 (8.7)    & non detectable \\
\hline
$u = 10$   & 4.2 (2.8)  & 7.8 (4.2)  & non detectable  \\
\hline 
$u  = 15$    & 2.0 (1.5) &  2.6 (2.0)   &  non detectable \\
\hline
$u = 20$    & non detectable  &  undetectable  & non detectable  \\

\hline
\end{tabular}
\end{center}

Table  6. The LHC cross sections for signal plus background in fb and significances (in brackets)  
for total energy $\sqrt{s} = 10~TeV$ and $M_{Z^`} = 1.8~TeV$. The background cross section is $\sigma_B = 0.28~fb$ 
and total integrated luminosity is $L_{tot} = 1~fb^{-1}$. 
\begin{center}
\begin{tabular}{|l| |l| |l| |l|}
\hline
 u-parameter  & model A   & model B & model C \\
\hline
 $u =0$  & 11 (5.6)  &  20 (8.0)    &  non detectable \\
\hline
$u =3$   &  22 (7.8)  & 19 (7.8)  & non detectable  \\
\hline
$u =5$  & 8.6 (4.8)  &  15 (6.3)    & non detectable \\
\hline
$u = 10$   & 3.2 (2.7)  & 6.4 (4.1)  & non detectable  \\
\hline 
$u  = 15$    & non detectable &  2.8 (1.7)   &  non detectable \\
\hline
$u = 20$    & non detectable  &  non detectable  & non detectable  \\

\hline
\end{tabular}
\end{center} 

Table  7. The LHC cross sections for signal plus background in fb and significances (in brackets)  
for total energy $\sqrt{s} = 14~TeV$ and $M_{Z^`} = 1.5~TeV$. The background cross section is $\sigma_B = 2.0~fb$ 
and total integrated luminosity is $L_{tot} = 100~fb^{-1}$. 
\begin{center}
\begin{tabular}{|l| |l| |l| |l|}
\hline
 u-parameter  & model A   & model B & model C \\
\hline
 $u =0$  & 82 (151)  &  154 (219)    &  22 (69) \\
\hline
$u =3$   &  78 (147)  & 138 (206)  & 5.2 (25)  \\
\hline
$u =5$  & 60 (126)  &  106 (175)    & 3.2 (17) \\
\hline
$u = 10$   & 26 (73)  & 48 (110)  & 1.1 (13) \\
\hline 
$u  = 15$    & 13 (41) &  22 (65)   &  0.48 (6.1) \\
\hline
$u = 20$    & 7.0 (25)  &  11 (37)  & .056 (3.7)  \\

\hline
\end{tabular}
\end{center} 

Table  8. The LHC cross sections for signal plus background in fb and significances (in brackets)  
for total energy $\sqrt{s} = 14~TeV$ and $M_{Z^`} = 2~TeV$. The background cross section is $\sigma_B = 0.44~fb$ 
and total integrated luminosity is $L_{tot} = 100~fb^{-1}$. 
\begin{center}
\begin{tabular}{|l| |l| |l| |l|}
\hline
 u-parameter  & model A   & model B & model C \\
\hline
 $u =0$  & 22 (80)  &  28 (110)    &  5.8 (36) \\
\hline
$u =3$   &  19 (75)  & 36 (107)  & 1.3 (13)  \\
\hline
$u =5$  &  15 (63)  &  26 (89)    & 0.72 (8.3) \\
\hline
$u = 10$   & 6.2 (37)  & 11 (53)  & 0.48 (5.9) \\
\hline 
$u  = 15$    & 2.8 (20) &  5.0 (31)   &  0.098 (3.2) \\
\hline
$u = 20$    & 1.6 (12)  &  2.4 (18)  & non detectable  \\

\hline
\end{tabular}
\end{center}

Table 9 .The LHC cross sections for signal plus background in fb and significances (in brackets)  
for total energy $\sqrt{s} = 14~TeV$ and $M_{Z^`} = 3~TeV$. The background cross section is $\sigma_B = 0.040~fb$ 
and total integrated luminosity is $L_{tot} = 100~fb^{-1}$. 
\begin{center}
\begin{tabular}{|l| |l| |l| |l|}
\hline
 u-parameter  & model A   & model B & model C \\
\hline
 $u =0$  & 2.0 (24)  &  3.6 (32)    &  0.54 (11) \\
\hline
$u =3$   & 1.9 (23)  & 3.4 (31)  & 0.12 (4.0)  \\
\hline
$u =5$  &  1.4 (20)  &  2.6 (28)    & 0.064 (2.4) \\
\hline
$u = 10$   & 0.54 (11)  & 1.0 (17)  & non detectable \\
\hline 
$u  = 15$    & 0.24 (5.6) &  0.46 (9.4)   &  non detectable \\
\hline
$u = 20$    & 0.13 (3.2)  &  0.22 (5.4)  & non detectable  \\

\hline
\end{tabular}
\end{center} 

Table 10. The LHC cross sections for signal plus background in fb and significances (in brackets)  
for total energy $\sqrt{s} = 14~TeV$ and $M_{Z^`} = 3.5~TeV$. The background cross section is $\sigma_B = 0.040~fb$ 
and total integrated luminosity is $L_{tot} = 100~fb^{-1}$. 
\begin{center}
\begin{tabular}{|l| |l| |l| }
\hline
 u-parameter  & model A   & model B  \\
\hline
 $u =0$  & 0.68 (14)  &  1.18 (19)     \\
\hline
$u =3$   & 0.60 (13)  & 1.12 (18)      \\
\hline
$u =5$  &  0.48  (11)  &  0.88 (16)    \\
\hline
$u = 10$   & 0.18 (6.4)  & 0.36 (10)   \\
\hline 
$u  = 15$    & 0.08 (3.6) &  0.14 (5.4)    \\
\hline
$u = 20$    & 0.042 (2.2)  &  0.070 (3.3)    \\

\hline
\end{tabular}
\end{center} 
$$  $$

As we can see from Tables 2-10 the LHC discovery potential depends 
rather strongly on 
the total decay width and it decreases with the increase of the 
total  decay width. The reason is trivial and it is 
due to the dilution factor $k = \frac{\Gamma_{tot}(u =o)}{\Gamma_{tot} (u=0) +\Gamma_{inv}(u)}$.
Besides the LHC discovery perspectives are the best for the model B and 
the 
worst for the model C.  The reason is that  the number of signal events 
is proportional to $\sigma Br$  and it is maximal for model B and minimal 
for model C (in model C the coupling constant of 
the $Z^`$ interaction with  quarks is small in comparison 
with  models A and B). 

Note that direct  Tevatron  
experimental bound on 
the mass of $Z^`$ boson depends on the $Z^`$ model and for $Z^`$ boson with standard couplings  
 $M_{Z^`} >  923~GeV$ \cite{20}. The existing experimental bounds on Stueckelberg $Z^`$ bosons 
were studied in ref.\cite{21}.
The very narrow Stueckelberg $Z^`$ boson can be discovered at Tevatron up to 
a mass  about $600 ~GeV $ with a total integrated luminosity of
$8 fb^{-1}$ \cite{21}.\footnote{LHC discovery perspectives for 
Stueckelberg $Z^`$ bosons were studied in ref.\cite{22}.}
For considered  $Z^`$ models 
with large invisible 
decay width  the  Tevaton   bound will be more  weak for wide $Z^`$ bosons.

To conclude in this paper we studied the  LHC discovery potential  for 
$Z^{'}$ models with 
continuously distributed mass.
One of  the possible LHC signatures for such models is the existence of   
broad  resonance structure in Drell-Yan reaction 
$$ pp \rightarrow Z^{'} + ... \rightarrow l^{+}l^{-} + ...$$. 
We made our estimates at the parton level 
and did not take into account detector level
\footnote{An account of detector effects will be given elsewhere 
\cite{23}.}.
Rough estimates show that detector effects are not very essential
 and will lead to the decrease of 
the significance by factor $0.8 - 0.9$. An account of K-factor 
partly compensates this decrease. 
The LHC discovery potential for wide  $Z`$ bosons is  more weak than 
for the case of narrow $Z`$ bosons
due to the dilution factor for $Br(Z^` \rightarrow l^+l^-) $. Nevertheless for 
some models  it is possible to detect  wide $ Z^`$ bosons for masses up to $3 - 4 ~TeV$ 
at final LHC energy $\sqrt{s} = 14~TeV$ and total integrated luminosity 
$L_{tot} = 100~fb^{-1}$. For the 2010-2011 LHC energy $\sqrt{s} = 7~TeV$ and 
total integrated luminosity $L_{tot} = 1~fb^{-1}$ it is possible to discover wide $Z^`$ boson 
with masses up to 1.2~TeV.

I am indebted to A.N.Toropin for useful discussions. 
This work was supported by the Grants RFBR  08-02-91007-CERN 
and 10-02-00468.

\end{document}